\documentclass[12pt]{article}
\usepackage{amsmath}
\usepackage{amssymb}
\usepackage{amsfonts}
\usepackage{graphics}
\usepackage [latin1]{inputenc}

\renewcommand{\theequation}{\arabic{section}.\arabic{equation}}

\title{Jacobi exceptional orthogonal polynomials for extended Scarf I potentials with position-dependent mass}
\author{C. Quesne\\ 
{\small\sl D\'epartement de Physique,  Universit\'e Libre de Bruxelles,} \\ 
{\small\sl Campus de la Plaine CP229, Boulevard~du Triomphe, B-1050 Brussels, Belgium}\\
{\small\sl E-mail: Christiane.Quesne@ulb.be}}
\date{ }
\begin{document}
\baselineskip=22pt plus 1pt minus 1pt
\maketitle

\begin{abstract}
We show that the Scarf I potential problem in a position-dependent mass background of the type $m(\alpha;x) = (1 + \alpha \sin x)^{-2}$, $0<\alpha<1$, can be solved by using a point canonical transformation mapping the corresponding Schr\" odinger equation onto that of the Scarf I potential with constant mass. The inverse point canonical transformation then provides some exactly-solvable rational extensions of the Scarf I potential with positive-dependent mass associated with $X_m$-Jacobi exceptional orthogonal polynomials of type I, II, or III. The Scarf I potential problem with position-dependent mass is shown to exhibit a deformed shape invariance property in a deformed supersymmetric framework. Such a property is also valid for extended potentials of type I and II. The results are illustrated with a simple example.
\end{abstract}

\section{Introduction}

A few years ago, new sets of orthogonal polynomials, generalizing the classical orthogonal polynomials of Hermite, Laguerre, and Jacobi \cite{olver}, were introduced and called exceptional orthogonal polynomials (EOPs) \cite{gomez09, gomez10}. They form orthogonal and complete polynomial sets with a positive weight function although there are some gaps in the sequence of their degrees in contrast with classical orthogonal polynomials \cite{gomez13, liaw, bonneux}. Soon after that, it was shown that they may occur in bound-state wavefunctions of rationally-extended well-known quantum potentials, constructed by Darboux transformations in the context of shape invariance in supersymmetric quantum mechanics \cite{cq08, cq09a, bagchi09, odake09, gomez12, odake11}.\par
%
%
Another topic that has recently received a lot of attention is that of position-dependent mass (PDM) in quantum systems because it has a lot of applications in several fields, ranging from the study of inhomogeneous materials to energy density many-body problems \cite{bastard, weisbuch, harrison, ring}. It is worth noting that the PDM presence in the Schr\"odinger equation may alternatively be interpreted as reflecting a curvature of the underlying space or a deformation of the canonical commutation relations \cite{cq04}.\par
%
%
A difficulty that occurs in such an approach comes from the momentum and mass operator noncommutativity, which gives rise to an ordering ambiguity in the kinetic energy representation. To cope with this difficulty, von Roos proposed a general two-parameter form of the kinetic energy operator \cite{vonroos}. A special case of the latter, which is very often used, is that of the Mustafa-Mazharimousavi ordering \cite{mustafa}.\par
%
%
In contrast with the case of standard Schr\"odinger equations, for with many exactly-solvable rational potentials connected with EOPs have been built (see, e.g., \cite{cq11, cq12, grandati, odake13, gomez14a, gomez14b, marquette}), for generalized Schr\"odinger equations there have been less attempts of constructing exactly-solvable rational potentials connected with EOPs. Among them, one may quote some extensions of the quantum oscillator and Kepler-Coulomb problems in curved space \cite{cq16}, those of an oscillator-shaped quantum well confined in a cavity between two infinite walls \cite{cq23}, those of a radial harmonic oscillator \cite{cq25}, as well as some extensions \cite{midya} of previously known potentials in a PDM background \cite{bagchi05a} and some families of asymmetric parabolic potentials associated with different kinds of PDM \cite{yadav}.\par
%
%
The aim of the present paper is to provide an additional example of exactly-solvable potentials with PDM, whose wavefunctions can be expressed in terms of $X_m$-Jacobi EOPs.\par
%
%
In section~2, we review the problem of a Scarf I potential in a $m(\alpha;x) = (1+ \alpha \sin x)^{-2}$ background and prove its exact solvability by mapping it onto that of a Scarf I potential in a constant mass background by means of a point canonical transformation (PCT). In section~3, we then make use of the known rational extensions for the latter problem connected with $X_m$-Jacobi EOPs to build similar extensions for the former one. The deformed supersymmetric properties of the resulting potentials are then discussed in section~4. A simple example is presented in section~5. Finally, section~6 contains the conclusion.\par
%
%
\section{\boldmath The Scarf I potential in a $m(\alpha;x) = (1 + \alpha \sin x)^{-2}$ background}

On using the Mustafa and Mazharimousavi ordering \cite{mustafa}, in units wherein $\hbar = 2m_0 = 1$, a PDM Schr\"odinger equation can be written as
\begin{equation}
  \biggl(- m^{-\frac{1}{4}}(\alpha;x) \frac{d}{dx} m^{-\frac{1}{2}}(\alpha;x) \frac{d}{dx} m^{-\frac{1}{4}}(\alpha;x)
  + V(x)\biggr) \psi^{(\alpha)}_n(x) = E^{(\alpha)}_n \psi^{(\alpha)}_n(x),  \label{eq:MM}
\end{equation}
where $m(\alpha;x)$ is the dimensionless form of the mass function $m_0 m(\alpha;x)$ and $V(x)$ is the potential. In the present paper, we assume that
\begin{equation}
  m(\alpha;x) = (1+ \alpha \sin x)^{-2}, \qquad 0<\alpha<1,
\end{equation}
and 
\begin{align}
  V(x) &= V(x;A,B) = [A(A-1) + B^2] \sec^2 x  - B(2A-1) \sec x \tan x,  \nonumber \\
  &\qquad A-1>B>0,\qquad -\frac{\pi}{2} < x < \frac{\pi}{2}, \label{eq:V}
\end{align}
is a Scarf I potential \cite{cooper}.\par
%
%
With such a choice of PDM, equation~(\ref{eq:MM}) may alternatively be interpreted as a deformed Schr\"odinger equation
\begin{equation}
  [\pi_x^2 + V(x;A,B)] \psi^{(\alpha)}_n(x;A,B) = E^{(\alpha)}_n(A,B) \psi^{(\alpha)}_n(x;A,B),  \label{eq:MM-bis}
\end{equation}
written in terms of a deformed momentum operator
\begin{equation}
  \pi_x = - {\rm i} \sqrt{f(\alpha;x)} \frac{d}{dx} \sqrt{f(\alpha;x)}, \qquad f(\alpha;x) = 1 + \alpha \sin x, \qquad
  0<\alpha<1.  \label{eq:f}
\end{equation}
\par
%
%
Equation (\ref{eq:MM-bis}) can be solved by mapping it onto a constant-mass Schr\"odinger equation
\begin{equation}
  \left(- \frac{d^2}{du^2} + U(u)\right) \phi_n(u) = \epsilon_n \phi_n(u)
\end{equation}
by a PCT \cite{bagchi04, cq09b}, i.e., by some changes of variable and of function
\begin{equation}
  u(\alpha;x) = a v(\alpha;x) + b, \qquad v(\alpha;x) = \int^x \frac{dx'}{f(\alpha;x')},  \label{eq:u}
\end{equation}
\begin{equation}
  \phi_n(u(\alpha;x)) \propto \sqrt{f(\alpha;x)} \psi^{(\alpha)}_n(x;A,B).
\end{equation}
The two potentials and their corresponding energy eigenvalues are related by
\begin{equation}
  V(x;A,B) = a^2 U(u) + c, \qquad E^{(\alpha)}_n(A,B) = a^2 \epsilon_n + c.  \label{eq:V-U}
\end{equation}
In (\ref{eq:u}) and (\ref{eq:V-U}), $a$, $b$, and $c$ denote three arbitrary real constants.\par
%
%
{}For the choice of $f(\alpha;x)$ made in (\ref{eq:f}), we obtain
\begin{equation}
  v(\alpha;x) = \frac{2}{\sqrt{1-\alpha^2}} \arctan \frac{\tan(x/2)+\alpha}{\sqrt{1-\alpha^2}}.
\end{equation}
Hence, on taking
\begin{equation}
  a = \sqrt{1-\alpha^2}, \qquad b = 2 \arctan \sqrt{\frac{1-\alpha}{1+\alpha}} - \frac{\pi}{2},  \label{eq:a-b} 
\end{equation}
we get
\begin{equation}
  u(\alpha;x) = 2 \arctan \frac{\tan(x/2)+\alpha}{\sqrt{1-\alpha^2}} + 2 \arctan \sqrt{\frac{1-\alpha}{1+\alpha}} 
  - \frac{\pi}{2},  \label{eq:u-bis}
\end{equation}
showing that $u$ varies in the interval $- \frac{\pi}{2} < u < \frac{\pi}{2}$. From (\ref{eq:u-bis}), it results that
\begin{equation}
  \tan u = \frac{1}{\sqrt{1-\alpha^2}} (\tan x + \alpha \sec x), \qquad \sec u = \frac{1}{\sqrt{1-\alpha^2}} (\sec x 
  + \alpha \tan x),
\end{equation}
or, conversely,
\begin{equation}
  \tan x = \frac{1}{\sqrt{1-\alpha^2}} (\tan u - \alpha \sec u), \qquad \sec x = \frac{1}{\sqrt{1-\alpha^2}} (\sec u
  - \alpha \tan u).
\end{equation}
\par
%
%
With such a choice of variable, the potential $V(x;A,B)$, defined in (\ref{eq:V}), becomes
\begin{equation}
  V(x;A,B) = (1-\alpha^2) U(u;A',B') + c,  \label{eq:V-U-bis}
\end{equation}
where $U(u;A',B')$ is another Scarf I potential
\begin{equation}
  U(u;A',B') = [A'(A'-1) + B^{\prime 2}] \sec^2u - B'(2A'-1) \sec u \tan u,  \label{eq:Scarf}
\end{equation}
whose parameters $A'$, $B'$ are defined in terms of $A$, $B$ by the equations
\begin{equation}
\begin{split}
  &A'(A'-1) + B^{\prime 2} = \frac{1}{(1-\alpha^2)^2}\{(1+\alpha^2) [A(A-1) + B^2] + 2\alpha B (2A-1)\}, \\
  &B'(2A'-1) = \frac{1}{(1-\alpha^2)^2} \{2\alpha [A(A-1) + B^2] + (1+\alpha^2) B(2A-1)\},
\end{split} \label{eq:parameters}
\end{equation} 
while the additional constant $c$ is given by
\begin{equation}
  c = - \frac{\alpha}{1-\alpha^2} \{\alpha[A(A-1)+B^2] + B(2A-1)\}.  \label{eq:c}
\end{equation}
\par
%
%
In order to solve (\ref{eq:parameters}) for $A'$ and $B'$, it is useful to introduce the combinations of parameters
\begin{equation}
  \Delta_1 = \left(\frac{1}{4} + \frac{(A-B)(A-B-1)}{(1+\alpha)^2}\right)^{1/2}, \qquad
  \Delta_2 = \left(\frac{1}{4} + \frac{(A+B)(A+B-1)}{(1-\alpha)^2}\right)^{1/2}.  \label{eq:Delta}
\end{equation}
Then $A'$ and $B'$ can be shown to be given by
\begin{equation}
  A' = \frac{1}{2}(\Delta_1 + \Delta_2 + 1), \qquad B' = \frac{1}{2}(\Delta_2 - \Delta_1),  \label{eq:A'B'}
\end{equation}
and, as a consequence of (\ref{eq:Delta}), they satisfy the condition $A'-1>B'>1$. It is worth observing that $c$ may be rewritten in terms of $A'$ and $B'$ as
\begin{equation}
  c = \alpha \{\alpha[A'(A'-1)+B^{\prime 2}] - B'(2A'-1)\}.  \label{eq:c-bis}
\end{equation}
\par
%
%
{}From the known eigenvalues
\begin{equation}
  \epsilon_n(A') = (A'+n)^2, \qquad n=0, 1, 2, \ldots,
\end{equation}
of the Scarf I potential $U(u;A',B')$ \cite{cooper}, and the constants $a$ and $c$ defined in (\ref{eq:a-b}) and (\ref{eq:c}) or (\ref{eq:c-bis}), respectively, we get for the eigenvalues (\ref{eq:V-U}) of the PDM Schr\"odinger equation
\begin{align}
  E^{(\alpha)}_n(A,B) &= (1-\alpha^2)n^2 + (1-\alpha^2)(\Delta_1+\Delta_2+1)n + \frac{1}{4}(1+\alpha)^2
      \Delta_1^2 + \frac{1}{4}(1-\alpha)^2 \Delta_2^2 \nonumber \\
  &\quad + \frac{1}{2}(1-\alpha^2)(\Delta_1\Delta_2+\Delta_1+\Delta_2) + \frac{1}{4}(1-2\alpha^2), \qquad
      n=0, 1, 2, \ldots,
\end{align}
provided the corresponding wavefunctions are normalizable on $\left(-\frac{\pi}{2}, \frac{\pi}{2}\right)$, which will indeed be the case.\par
%
%
Such wavefunctions can be expressed as
\begin{equation}
  \psi^{(\alpha)}_n(x;A,B) = \lambda [f(\alpha;x)]^{-1/2} \phi_n(u;A',B'),  \label{eq:psi-phi}
\end{equation}
where $\phi_n(u;A',B')$ are the well-known Scarf I wavefunctions \cite{cq09a}
\begin{equation}
\begin{split}
  \phi_n(u;A',B') &= {\cal N}^{(A',B')}_n (1 - \sin u)^{\frac{1}{2}(A'-B')} (1+ \sin u)^{\frac{1}{2}(A'+B')} \\
  &\quad \times P_n^{(A'-B'-\frac{1}{2}, A'+B'-\frac{1}{2})}(\sin u), \\
  {\cal N}^{(A',B')}_n &= \left(\frac{(2A'+2n) n! \Gamma(2A'+n)}{2^{2A'} \Gamma\left(A'-B'+n+\frac{1}{2}\right)
       \Gamma\left(A'+B'+n+\frac{1}{2}\right)}\right)^{1/2},  \label{eq:phi}
\end{split}
\end{equation}
written in terms of Jacobi polynomials, and $\lambda$ is the constant $(1-\alpha^2)^{1/4}$, arising from the change of normalization from $u \in \left(-\frac{\pi}{2}, \frac{\pi}{2}\right)$ to $x \in \left(-\frac{\pi}{2}, \frac{\pi}{2}\right)$. After some straightforward calculations, one gets
\begin{align}
  \psi^{(\alpha)}_n(x;A,B) &= {\cal N}^{(\alpha)}_{n,A,B} [f(\alpha;x)]^{-\frac{1}{2}(\Delta_1+\Delta_2+2)}
          (1-\sin x)^{\frac{1}{2}\left(\Delta_1+\frac{1}{2}\right)} \nonumber \\
  &\quad \times (1+ \sin x)^{\frac{1}{2}\left(\Delta_2+\frac{1}{2}\right)} P_n^{(\Delta_1,\Delta_2)}(t), \qquad
          t = \frac{\alpha+\sin x}{1+\alpha \sin x}, \label{eq:psi} 
\end{align}
with
\begin{equation}
  {\cal N}^{(\alpha)}_{n,A,B} = \left(\frac{(1-\alpha)^{\Delta_1+1} (1+\alpha)^{\Delta_2+1} (\Delta_1 + \Delta_2 + 1 
  +2n) n! \Gamma(\Delta_1 + \Delta_2 + 1 +n)} {2^{\Delta_1+\Delta_2+1} \Gamma(\Delta_1+n+1)
  \Gamma(\Delta_2+n+1)}\right)^{1/2},       
\end{equation}
and $n=0$, 1, 2, \ldots . \par
%
%
\section{\boldmath Rational extensions of the Scarf I potential in a $m(\alpha;x) = (1+\alpha \sin x)^{-2}$ background}

\setcounter{equation}{0}

Rational extensions of the Scarf I potential connected with Jacobi EOPs and a constant mass have been constructed as partners of conventional potentials in the framework of SUSYQM \cite{cq08, cq09a, odake09}. We will consider here more specifically the case of one-step SUSYQM, leading to $X_m$-Jacobi polynomials. By applying the inverse of the PCT presented in section~2, we will build from them rational extensions of the Scarf I potential in the $m(\alpha;x)$ background and study some properties of the latter.\par
%
%
The rational extensions of the Scarf I potential $U(u;A',B')$ belong to three different types I, II, and III, according to the kind of seed function that is used to construct the partner. Such seed functions and their corresponding energies can be written as
\begin{equation}
\begin{split}
  &\varphi^{\rm I}_m(u;A',B') = \phi_m(u; B'+\tfrac{1}{2},A'-\tfrac{1}{2}), \qquad e^{\rm I}_m(A',B') = 
        (B'+m+\tfrac{1}{2})^2, \\
  & \quad A'-B' > m + \tfrac{1}{2}, \\
  &\varphi^{\rm II}_m(u;A',B') = \phi_m(u; \tfrac{1}{2}-B', \tfrac{1}{2}-A'), \qquad e^{\rm II}_m(A',B') = 
        (B'-m-\tfrac{1}{2})^2, \\
  & \quad A'+B' > m + \tfrac{1}{2}, \\
  &\varphi^{\rm III}_m(u;A',B') = \phi_m(u; 1-A',-B'), \qquad e^{\rm III}_m(A',B') = (A'-m-1)^2,
       \\ 
  & \quad A'+B', A'-B' > m + \tfrac{1}{2}, \qquad m \text{\ even}, 
\end{split}  \label{eq:varphi}
\end{equation}
where
\begin{equation}
  \phi_m(u;A',B') \propto (1-\sin u)^{\frac{1}{2}(A'-B')} (1+\sin u)^{\frac{1}{2}(A'+B')} P_m^{(A'-B'-\frac{1}{2},
  A'+B'-\frac{1}{2})}(\sin u),
\end{equation}
in accordance with (\ref{eq:phi}).\par
%
%
To obtain for the partner some rationally-extended Scarf I potential with given $A'$ and $B'$ parameters, we have to start from a conventional potential with different parameters $\bar{A}'$, $\bar{B}'$, which depend on the type considered. The results read
\begin{equation}
\begin{split}
  U_0(u) &= U(u;\bar{A}',\bar{B}'), \\
  U_1(u) &= U(u;\bar{A}',\bar{B}') - 2 \frac{d^2}{du^2} \log \varphi_m^i(u;\bar{A}',\bar{B}') \\
  &= U^{(m)}_{\rm ext}(u;A',B') = U(u;A',B') + U^{(m)}_{\rm rat}(u;A',B'),
\end{split}  \label{eq:U_0-U_1}
\end{equation} 
where
\begin{equation}
  U^{(m)}_{\rm rat}(u;A',B') = 2\left\{z \frac{\dot{g}_m^{(A',B')}}{g_m^{(A',B')}} - (1-z^2)\left[
  \frac{\ddot{g}_m^{(A',B')}}{g_m^{(A',B')}} - \left(\frac{\dot{g}_m^{(A',B')}}{g_m^{(A',B')}}\right)^2\right]
  \right\}, \qquad z = \sin u.
\end{equation}
Here a dot denotes a derivative with respect to $z$, and we have for the three different types
\begin{equation}
\begin{split}
  \text{(I)\ } &\bar{A}'=A', \quad \bar{B}'=B'-1, \quad g_m^{(A',B')}(z) = P_m^{\left(-A'+B'-\frac{1}{2}, 
       A'+B'-\frac{3}{2}\right)}(z), \\
  & \quad A'-B' > m-\tfrac{1}{2}, \\
  \text{(II)\ } &\bar{A}'=A', \quad \bar{B}'=B'+1, \quad g_m^{(A',B')}(z) = P_m^{\left(A'-B'-\frac{3}{2}, 
       -A'-B'-\frac{1}{2}\right)}(z), \\
  & \quad A'+B' > m-\tfrac{1}{2},  \\  
  \text{(III)\ } &\bar{A}'=A'+1, \quad \bar{B}'=B', \quad g_m^{(A',B')}(z) = P_m^{\left(-A'+B'-\frac{1}{2}, 
       -A'-B'-\frac{1}{2}\right)}(z), \\
  & \quad A'+B', A'-B' > m -\tfrac{1}{2}, \quad m \text{\ even}.     
\end{split}  \label{eq:g}
\end{equation}
It is worth noting that on going from $(A', B')$ to $(\bar{A}', \bar{B}')$, $\Delta_1$ and $\Delta_2$ in (\ref{eq:Delta}) are changed into
\begin{equation}
  (\bar{\Delta}_1, \bar{\Delta_2}) = 
    \begin{cases}
       (\Delta_1+1, \Delta_2-1) & \text{for type I}, \\
       (\Delta_1-1, \Delta_2+1) & \text{for type II}, \\
       (\Delta_1+1, \Delta_2+1) & \text{for type III}.    
    \end{cases} \label{eq:Delta-bar}
\end{equation}
\par
%
%
{}For types I and II, the two partners $U_0(u)$ and $U_1(u)$ are strictly isospectral, so that the spectrum of the latter is given by
\begin{equation}
  \epsilon_n^{(\rm ext)}(A',B') = (A'+n)^2, \qquad n=0, 1, 2, \ldots, \qquad \text{for type I or II},
\end{equation}
whereas, for type III, the inverse of the seed function being normalizable, we get an extra bound state below the spectrum  of $U_0(u)$, hence
\begin{equation}
  \epsilon_n^{(\rm ext)}(A',B') = (A'+1+n)^2, \qquad n=-m-1, 0, 1, 2, \ldots, \qquad \text{for type III}.
\end{equation}
\par
%
%
The wavefunctions of $U_1(u)$ corresponding to $n=0$, 1, 2,~\ldots, can be obtained from
\begin{equation}
  \phi^{(\rm ext}_n(u;A',B') \propto \frac{{\cal W}(\varphi^i_m(u;\bar{A}',\bar{B}'), \phi_n(u;\bar{A}',\bar{B}') 
  \mid u)}{\varphi^i_m(u;\bar{A}',\bar{B}')}, \qquad i = \text{{\rm I}, {\rm II}, or {\rm III}}, 
\end{equation}
where ${\cal W}(\varphi^i_m, \phi_n \mid u)$ denotes the Wronskian of the functions $\varphi^i_m$ and $\phi_n$ \cite{muir}. From Eqs.~(\ref{eq:phi}), (\ref{eq:varphi}), (\ref{eq:g}), and standard properties of Wronskians, they can be written as
\begin{equation}
  \phi^{\rm ext}_n(u;A',B') \propto \frac{\phi_0(u;A',B')}{g^{(A',B')}_m(z)} Q^{(m)}_n(z;A',B'), \qquad 
  n=0,1,2, \ldots,  \label{eq:phi-ext}
\end{equation}
where $Q^{(m)}_n(z;A',B')$ is given by
\begin{align}
  Q^{(m)}_n(z;A',B') &= - \left(A'-B'+\frac{1}{2}\right) g^{(A',B')}_m(z) P_n^{(A'-B'+\frac{1}{2},A'+B'-\frac{3}{2})} 
       (z)\nonumber \\
  &\quad +\frac{1}{2}(1-z) \biggl[(n+2A') g^{(A',B')}_m(z) P^{(A'-B'+\frac{3}{2}, A'+B'-\frac{1}{2})}_{n-1}(z)
       \nonumber \\
  &\quad - (m+2B'-1) g^{(A',B'+1)}_{m-1}(z) P^{(A'-B'+\frac{1}{2},A'+B4-\frac{3}{2})}_n(z)\biggr],
\end{align}
\begin{align}
  Q^{(m)}_n(z;A',B') &= \left(A'+B'+\frac{1}{2}\right) g^{(A',B')}_m(z) P^{(A'-B'-\frac{3}{2}, A'+B'+\frac{1}{2})}_n(z)
      \nonumber \\
  &\quad + \frac{1}{2}(1+z) \biggl[(n+2A') g^{(A',B')}_m(z) P^{(A'-B'-\frac{1}{2}, A'+B'+\frac{3}{2})}_{n-1}(z)
      \nonumber \\
  &\quad -(m-2B'-1) g^{(A',B'-1)}_{m-1}(z) P^{(A'-B'-\frac{3}{2}, A'+B'+\frac{1}{2})}_n(z)\biggr],
\end{align}
\begin{align}
  Q^{(m)}_n(z;A',B') &= [2B' - (2A'+1)z] g^{(A',B')}_m(z) P^{(A'-B'+\frac{1}{2}, A'+B'+\frac{1}{2})}_n(z) \nonumber \\
  &\quad +\frac{1}{2}(1-z^2) \biggl[(n+2A'+2) g^{(A',B')}_m(z)  P^{(A'-B'+\frac{3}{2},A'+B'+\frac{3}{2})}_{n-1}(z)
       \nonumber \\
  &\quad - (m-2A') g^{(A'-1,B')}_{m-1}(z) P^{(A'-B'+\frac{1}{2},A'+B'+\frac{1}{2})}_n(z)\biggr],  \label{eq:Q}
\end{align}
for types I, II, and III, respectively. In the first two cases, it is a $(m+n)$th-degree polynomial in $z$, while for type III, it is a $(m+n+1)$th-degree polynomial in $z$. In the latter case, Eq.~(\ref{eq:phi-ext}) is also valid for $n=-m-1$ and $Q^{(m)}_{-m-1}(z;A',B') = 1$.\par
%
%
Due to the orthogonality properties of bound-state wavefunctions, in all three cases, the polynomials $Q^{(m)}_n(z;A',B')$ constitute families of orthogonal polynomials on $(-1,+1)$ with respect to the measure $(1-z)^{A'-B'-\frac{1}{2}} (1+z)^{A'+B'-\frac{1}{2}} \left(g_m^{(A',B')}(z)\right)^{-2}$. From the absence of scattering states, it results that these families also form complete sets and are therefore EOP families.\par
%
%
Let us now consider the inverse of the PCT mapping the Scarf I potential with PDM onto another one with constant mass, as defined in Section~2, to determine the seed functions $\chi^{\rm i}_m(x;A,B)$, ${\rm i} = {\rm I}, {\rm II}$, or {\rm III}, associated with $\phi_m^{\rm i}(u;A',B')$, ${\rm i} = {\rm I}, {\rm II}$, or {\rm III}. From (\ref{eq:varphi}), it is clear that $(A',B')$ is replaced by $(B'+\frac{1}{2}, A'-\frac{1}{2})$, $(\frac{1}{2}-B', \frac{1}{2}-A')$, or $(1-A', -B')$ in the definition of the seed function, respectively. From (\ref{eq:A'B'}), such a transformation can be made by changing $(\Delta_1,\Delta_2)$ into $(-\Delta_1,\Delta_2)$, $(\Delta_1,-\Delta_2)$, or $(-\Delta_1,-\Delta_2)$, respectively. From (\ref{eq:psi}), we therefore get
\begin{equation}
\begin{split}
  \chi_m^{\rm I}(x;A,B) &= f^{-\frac{1}{2}(-\Delta_1+\Delta_2+2)} (1-\sin x)^{\frac{1}{2}(-\Delta_1+\frac{1}{2})}
       (1+\sin x)^{\frac{1}{2}(\Delta_2+\frac{1}{2})} \\
  &\quad \times P_m^{(-\Delta_1,\Delta_2)}(t), \qquad m<\Delta_1, \\
  \chi_m^{\rm II}(x;A,B) &= f^{-\frac{1}{2}(\Delta_1-\Delta_2+2)} (1-\sin x)^{\frac{1}{2}(\Delta_1+\frac{1}{2})}
        (1+\sin x)^{\frac{1}{2}(-\Delta_2+\frac{1}{2})} \\
  &\quad \times P_m^{(\Delta_1,-\Delta_2)}(t), \qquad m<\Delta_2, \\
  \chi_m^{\rm III}(x;A,B) &= f^{-\frac{1}{2}(-\Delta_1-\Delta_2+2)} (1-\sin x)^{\frac{1}{2}(-\Delta_1+\frac{1}{2})}
        (1+\sin x)^{\frac{1}{2}(-\Delta_2+\frac{1}{2})} \\
  &\quad\times P_m^{(-\Delta_1,-\Delta_2)}(t), \qquad m<\Delta_1, \Delta_2, \qquad \text{$m$ even},
\end{split} \label{eq:chi}
\end{equation}
with corresponding energies
\begin{equation}
\begin{split}
  {\cal E}_m^{\rm I}(A,B) &= (1-\alpha^2)m^2 + (1-\alpha^2)(-\Delta_1+\Delta_2+1)m + \frac{1}{4}(1+\alpha)^2 
       \Delta_1^2  \\
   &\quad + \frac{1}{4}(1-\alpha)^2 \Delta_2^2 + \frac{1}{2}(1-\alpha^2)(-\Delta_1\Delta_2-\Delta_1+\Delta_2) 
        + \frac{1}{4}(1-2\alpha^2), \\
   {\cal E}_m^{\rm II}(A,B) &= (1-\alpha^2)m^2 + (1-\alpha^2)(\Delta_1-\Delta_2+1)m + \frac{1}{4}(1+\alpha)^2 
       \Delta_1^2  \\
   &\quad + \frac{1}{4}(1-\alpha)^2 \Delta_2^2 + \frac{1}{2}(1-\alpha^2)(-\Delta_1\Delta_2+\Delta_1-\Delta_2) 
       + \frac{1}{4}(1-2\alpha^2), \\
  {\cal E}_m^{\rm III}(A,B) &= (1-\alpha^2)m^2 + (1-\alpha^2)(-\Delta_1-\Delta_2+1)m + \frac{1}{4}(1+\alpha)^2 
       \Delta_1^2  \\
   &\quad + \frac{1}{4}(1-\alpha)^2 \Delta_2^2+ \frac{1}{2}(1-\alpha^2)(\Delta_1\Delta_2-\Delta_1-\Delta_2) 
       + \frac{1}{4}(1-2\alpha^2).
\end{split} \label{eq:cal-E}
\end{equation}
\par
%
%
{}Furthermore, the pair of potentials $U_0(u)$ and $U_1(u)$, defined in (\ref{eq:U_0-U_1})--(\ref{eq:g}), gives rise to partners
\begin{equation}
\begin{split}
  V_0(x) &= V(x;\bar{A},\bar{B}), \\
  V_1(x) &= V^{(m)}_{\rm ext}(x;A,B) + \gamma = V(x;A,B)) + V^{(m)}_{\rm rat}(x;A,B) + \gamma,  
  \label{eq:V_0-V_1}
\end{split}
\end{equation}
where
\begin{equation}
  V^{(m)}_{\rm rat}(x;A,B) = 2(1-\alpha^2) \left\{t \frac{\dot{p}_m^{(\Delta_1,\Delta_2)}}{p_m^{(\Delta_1,\Delta_2)}}
  - (1-t^2) \left[\frac{\ddot{p}_m^{(\Delta_1,\Delta_2)}}{p_m^{(\Delta_1,\Delta_2)}} - 
  \left(\frac{\dot{p}_m^{(\Delta_1,\Delta_2)}}{p_m^{(\Delta_1,\Delta_2)}}\right)^2\right]\right\}, \label{eq:Vrat1}
\end{equation}
and
\begin{equation}
  p_m^{(\Delta_1,\Delta_2)}(t) = 
     \begin{cases}
        P_m^{(-\Delta_1-1,\Delta_2-1)}(t) & \text{for type I}, \\
        P_m^{(\Delta_1-1,-\Delta_2-1)}(t) & \text{for type II}, \\
        P_m^{(-\Delta_1-1,-\Delta_2-1)}(t) & \text{for type III},     
     \end{cases} \label{eq:p_m}
\end{equation}
resulting from (\ref{eq:V-U-bis}) and (\ref{eq:A'B'}), with $t = (\alpha+\sin x)/(1+\alpha \sin x)$ and a dot denoting a derivative with respect to $t$. In (\ref{eq:V_0-V_1}), $\bar{A}$ and $\bar{B}$ are parameters associated with $\bar{A}'$ and $\bar{B}'$, defined in ({\ref{eq:g}). Their explicit form in terms of $A$, $B$, and $\alpha$ is determined in the appendix. The presence of the additive constant $\gamma$ in (\ref{eq:V_0-V_1}) is due to the dependence of $c$ on $A'$ and $B'$ (see (\ref{eq:c-bis})) and to the fact that the latter assume different values for the two partners. So for type I, for instance, the PCT changes $U(u;A',B'-1)$ into $V(x;\bar{A},\bar{B}) = (1-\alpha^2) U(u;A',B'-1) + \alpha\{\alpha[A'(A'-1)+(B'-1)^2] - (B'-1)(2A'-1)\}$, whereas $U^{(m)}_{\rm ext}(x;A',B')$ is modified into $V^{(m)}_{\rm ext}(x;A,B) = (1-\alpha^2)U^{(m}_{\rm ext}(x;A',B') + \alpha\{\alpha [A'(A'-1)+B^{\prime 2}] - B'(2A'-1)\}$. Since $U(u;A',B'-1)$ and $U^{(m)}_{\rm ext}(u;A',B')$ are isospectral, to get the same property for the image potentials, we have to consider $V(x;\bar{A},\bar{B})$ and 
$V^{(m)}_{\rm ext}(x;A,B) - \alpha\{\alpha [A'(A'-1)+B^{\prime 2}] - B'(2A'-1)\} + \alpha\{\alpha[A'(A'-1)+(B'-1)^2] - (B'-1)(2A'-1)\} = V^{(m)}_{\rm ext}(x;A,B) + \alpha(2A'-2\alpha B'+\alpha-1) = V^{(m)}_{\rm ext}(x;A,B) + \alpha[(1+\alpha)\Delta_1 + (1-\alpha)\Delta_2 + \alpha]$. A similar reasoning applies to the remaining two cases. Hence $\gamma$ in (\ref{eq:V_0-V_1}) is given by
\begin{equation}
  \gamma = 
     \begin{cases}
         \alpha[(1+\alpha)\Delta_1 + (1-\alpha)\Delta_2 + \alpha] &\text{for type I}, \\
         \alpha[-(1+\alpha)\Delta_1 - (1-\alpha)\Delta_2 + \alpha] &\text{fpr type II}, \\
         \alpha[(1+\alpha)\Delta_1 - (1-\alpha)\Delta_2 + \alpha] &\text{for type III}.  
      \end{cases}
\end{equation}
\par
%
%
{}For type I or II, the spectra of $V_0(x)$ and $V_1(x)$ are given by $E^{(\alpha)}_n(\bar{A},\bar{B})$, $n=0, 1, 2, \ldots$. Hence, the spectrum of $V^{(m)}_{\rm ext}(x;A,B)$, given by $E^{(\alpha)}_n(\bar{A},\bar{B}) - \gamma$, turns out to be the same as that of $V(x;A,B)$, i.e.,
\begin{align}
  E^{(\rm ext)}_n(A,B) &= (1-\alpha^2) n^2 + (1-\alpha^2) (\Delta_1+\Delta_2+1) n + \frac{1}{4}(1+\alpha)^2
         \Delta_1^2 + \frac{1}{4}(1-\alpha)^2 \Delta_2^2 \nonumber \\
  &\quad + \frac{1}{2}(1-\alpha^2) (\Delta_1\Delta_2+\Delta_1+\Delta_2) + \frac{1}{4}(1-2\alpha^2), \qquad
         n=0, 1, 2, \ldots,  \label{eq:E-ext-n}
\end{align}
for type I or II. On the other hand, we get
\begin{align}
  E^{(\rm ext)}_n(A,B) &= (1-\alpha^2) n^2 + (1-\alpha^2) (\Delta_1+\Delta_2+3) n + \frac{1}{4}(1+\alpha)^2
         \Delta_1^2 + \frac{1}{4}(1-\alpha)^2 \Delta_2^2 \nonumber \\
  &\quad + \frac{1}{2}(1-\alpha^2) (\Delta_1\Delta_2+3\Delta_1+3\Delta_2) + \frac{9}{4} - \frac{5}{2}\alpha^2,        
         \nonumber \\
         &\quad n=-m-1,0, 1, 2, \ldots,
\end{align}
for type III.\par
%
%
The corresponding wavefunctions, obtained from Eqs.~(\ref{eq:psi-phi}) and (\ref{eq:phi-ext})--(\ref{eq:Q}), read
\begin{equation}
  \psi_n^{(\rm ext)}(x;A,B) \propto \frac{\psi^{(\alpha)}_0(x;A,B)}{p_m^{(\Delta_1,\Delta_2)}(t)} 
  Q^{(m)}_n(t;A,B), \qquad t = \frac{\alpha+\sin x}{1+\alpha\sin x}, \label{eq:wf-ext-1}
\end{equation}
where the polynomials $Q^{(m)}_n(t;A,B)$ can be written as
\begin{align}
  &Q^{(m)}_n(t;A,B) = -(\Delta_1+1) p_m^{(\Delta_1,\Delta_2)}(t) P_n^{(\Delta_1+1,\Delta_2-1)}(t) \nonumber \\
  &\quad + \frac{1}{2}(1-t) \biggl[(n+\Delta_1+\Delta_2+1) p_m^{(\Delta_1,\Delta_2)}(t) P_{n-1}^{(\Delta_1+2,
        \Delta_2)}(t) \nonumber \\
  &\quad - (m+\Delta_2-\Delta_1-1) p_{m-1}^{(\Delta_1-1,\Delta_2+1)}(t) P_n^{(\Delta_1+1,\Delta_2-1)}(t)\biggr], 
         \qquad n=0,1,2,\ldots,  \label{eq:wf-ext-1a}
\end{align}
\begin{align}
  &Q^{(m)}_n(t;A,B) = (\Delta_2+1) p_m^{(\Delta_1,\Delta_2)}(t) P_n^{(\Delta_1-1,\Delta_2+1)}(t)
        \nonumber \\
  &\quad+\frac{1}{2}(1+t)\biggl[(n+\Delta_1+\Delta_2+1) p_m^{(\Delta_1,\Delta_2)}(t) P_{n-1}^{(\Delta_1,
        \Delta_2+2)}(t)  \nonumber \\
  &\quad-(m+\Delta_1-\Delta_2-1) p_{m-1}^{(\Delta_1+1,\Delta_2-1)}(t) P_n^{(\Delta_1-1,\Delta_2+1)}(t)
       \biggr], 
        \qquad n=0,1,2,\ldots,  \label{eq:wf-ext-1b}
\end{align}
and
\begin{align}
  &Q^{(m)}_n(t;A,B) = [\Delta_2-\Delta_1 - (\Delta_1+\Delta_2+2)t] p_m^{(\Delta_1,\Delta_2)}(t)  P_n^{(\Delta_1+1,
       \Delta_2+1)}(t) \nonumber \\
  &\quad+ \frac{1}{2}(1-t^2) \biggl[(n+\Delta_1+\Delta_2+3) p_m^{(\Delta_1,\Delta_2)}(t) P_{n-1}^{(\Delta_1+2,
       \Delta_2+2)}(t) \nonumber \\
  &\quad-(m-\Delta_1-\Delta_2-1) p_{m-1}^{(\Delta_1-1,\Delta_2-1)}(t) P_n^{(\Delta_1+1,\Delta_2+1)}(t)\biggr], 
        \qquad n=0,1,2,\ldots, \nonumber\\
  &Q^{(m)}_{-m-1}(t;A,B) = 1,  \label{eq:wf-ext-2}
\end{align}
for types I, II, and III, respectively. They constitute orthogonal and complete families of polynomials (i.e., EOPs) on $(-1,+1)$ with respect to the measure $(1-t)^{\Delta_1} (1+t)^{\Delta_2} \left(p_m^{(\Delta_1,\Delta_2)}(t)\right)^{-2}$.\par
%
%
\section{Deformed supersymmetric properties}

\setcounter{equation}{0}

It is well known that the Scarf I potential $U(u;A',B')$, defined in (\ref{eq:Scarf}), is a shape invariant potential in the context of SUSYQM \cite{cooper}. As a result, the transformed potential $V(x;A,B)$ of Eq.~(\ref{eq:V}) is deformed shape invariant in the context of deformed SUSYQM \cite{bagchi05b}.\par
%
%
This means that on taking the ground-state wavefunction $\psi^{(\alpha)}_0(x;A,B) = {\cal N}^{(\alpha)}_{0,A,B} [f(\alpha;x)]^{-\frac{1}{2}(\Delta_1+\Delta_2+2)} (1-\sin x)^{\frac{1}{2}\left(\Delta_1+\frac{1}{2}\right)} (1+ \sin x)^{\frac{1}{2}\left(\Delta_2+\frac{1}{2}\right)}$ of the deformed Schr\"odinger equation (\ref{eq:MM-bis}) as the seed function, the corresponding superpotential
\begin{equation}
  W_{\alpha}(x;A,B) = - f \frac{d}{dx} \log \psi^{(\alpha)}_0(x;A,B) - \frac{1}{2} f',
\end{equation}
where a dash denotes a derivative with respect to $x$, can be written as
\begin{equation}
  W_{\alpha}(x;A,B) = \lambda \tan x + \mu \sec x,
\end{equation}
with 
\begin{equation}
  \lambda = \frac{1}{2}[\Delta_1+\Delta_2+1 + \alpha(\Delta_1-\Delta_2)], \qquad \mu = \frac{1}{2}[\Delta_1
  -\Delta_2 + \alpha(\Delta_1+\Delta_2+1)].
\end{equation}
It allows us to consider a pair of first-order differential operators
\begin{equation}
  A^{\pm}_{\alpha}(A,B) = \mp \sqrt{f} \frac{d}{dx} \sqrt{f} + W_{\alpha}(x;A,B),
\end{equation}
in terms of which the starting deformed Hamiltonian $H_0 = \pi_x^2 + V_0(x)$, $V_0(x) = V(x;A,B)$, can be written as
\begin{equation}
  H_0 = A^+_{\alpha}(A,B) A^-_{\alpha}(A,B) + E^{(\alpha)}_0(A,B).
\end{equation}
\par
%
%
The partner of $H_0$ is
\begin{align}
  &H_1 = A^-_{\alpha}(A,B) A^+_{\alpha}(A,B) + E^{(\alpha)}_0(A,B) = \pi_x^2 + V_1(x), \nonumber \\
  & V_1(x) = V_0(x) + 2f \frac{d}{dx} W_{\alpha}(x;A,B).
\end{align}
On writing $A= A^{(0)}$, $B=B^{(0)}$, so that $V_0(x) = V(x;A^{(0)},B^{(0)})$ and $\Delta_1=\Delta_1^{(0)}$, $\Delta_2=\Delta_2^{(0)}$, one finds that
\begin{equation}
  V_1(x) = V(x;A^{(1)},B^{(1)}) - \alpha \left[(1+\alpha)\Delta_1^{(0)} - (1-\alpha) \Delta_2^{(0)} + \alpha\right],
  \label{eq:V_1}
\end{equation}
where
\begin{equation}
\begin{split}
  A^{(1)}(A^{(1)}-1) + (B^{(1)})^2 &= A^{(0)}(A^{(0)}-1) + (B^{(0)})^2 + (1+\alpha)^2 \Delta_1^{(0)}
      + (1-\alpha)^2 \Delta_2^{(0)}\\
  &\quad + 1 + \alpha^2, \\
  - B^{(1)} (2A^{(1)}-1) &= - B^{(0)} (2A^{(0)}-1) + (1+\alpha)^2 \Delta_1^{(0)} - (1-\alpha)^2 \Delta_2^{(0)}
      + 2\alpha,
\end{split}. \label{eq:A1-B1}
\end{equation}
and $A^{(1)}$, $B^{(1)}$ correspond to  $\Delta_1^{(1)} = \Delta_1^{(0)}+1$, $\Delta_2^{(1)} = \Delta_2^{(0)}+1$. Hence, the potential $V_0(x) = V(x;A,B)$ exhibits a deformed shape invariance property: up tp some additive constant, its partner $V_1(x)$ is similar in shape and differs only in the parameters $A^{(1)}$, $B^{(1)}$ that appear in it. The explicit form of $A^{(1)}$ and $B^{(1)}$ in terms of $A^{(0)}$ and $B^{(0)}$ is determined in the appendix.\par
%
%
Such a deformed shape invariance property enables us to construct a hierarchy of Hamiltonians
\begin{equation}
  H_i = A^+_{\alpha}(A^{(i)}, B^{(i)}) A^-_{\alpha}(A^{(i)}, B^{(i)}) + \sum_{j=0}^i \varepsilon_j, \qquad i=0, 1,2, \ldots,
\end{equation}
where
\begin{equation}
\begin{split}
  & A^{\pm}_{\alpha}(A^{(i)}, B^{(i)}) = \mp \sqrt{f} \frac{d}{dx} \sqrt{f} + W_{\alpha}(x; A^{(i)}, B^{(i)}), \\
  & \varepsilon_0 = E^{(\alpha)}_0(A, B), \\
  & \varepsilon_i = (1-\alpha^2) (\Delta_1 + \Delta_2 + 2i) = 2\lambda - 2\alpha \mu + (1-\alpha^2)(2i-1), \qquad i=1, 2, 
       \ldots,
\end{split}
\end{equation}
and
\begin{equation}
  W_{\alpha}(x; A^{(i)}, B^{(i)}) = (\lambda+i) \tan x + (\mu+i \alpha) \sec x.
\end{equation}
Such Hamiltonians are associated with a set of potentials $V_i(x)$, $i=0$, 1, 2, \ldots, in such a way that
\begin{equation}
  H_i = \pi_x^2 + V_i(x),
\end{equation}
where
\begin{equation}
  V_i(x) = V(x; A^{(i)}, B^{(i)}) - i \alpha [\Delta_1 - \Delta_2 + \alpha(\Delta_1 + \Delta_2 + i)].
\end{equation}
\par
%
%
One may also write
\begin{equation}
  H_{i+1} = A^-_{\alpha}(A^{(i)}, B^{(i)}) A^+_{\alpha}(A^{(i)}, B^{(i)}) + \sum_{j=0}^i \varepsilon_j.
\end{equation}
In other words, the first-order operators $A^{\pm}_{\alpha}(A^{(i)}, B^{(i)})$ fulfil a deformed shape invariance condition
\begin{align}
  & A^-_{\alpha}(A^{(i)}, B^{(i)}) A^+_{\alpha}(A^{(i)}, B^{(i)}) = A^+_{\alpha}(A^{(i+1)}, B^{(i+1)})
      A^-_{\alpha}(A^{(i+1)}, B^{(i+1)}) + \varepsilon_{i+1}, \nonumber\\
  & \qquad i = 0, 1, 2, \ldots.
\end{align}
\par
%
%
{}Finally, one may note that the energy eigenvalues of the starting Hamiltonian $H_0$ can be directly found from the $\varepsilon_i$'s as
\begin{equation}
  E^{(\alpha)}_n(A, B) = \sum_{i=0}^n \varepsilon_i.
\end{equation}
\par
%
%
On considering next the rationally-extended Scarf I potentials with PDM, obtained in (\ref{eq:V_0-V_1}) by using the PCT, let us first review the deformed supersymmetric framework wherein they may be alternatively derived.
%
%
{}For the potentials (\ref{eq:V_0-V_1}), let us consider a superpotential
\begin{equation}
  W^{(m)}_{\alpha}(x; \bar{A}, \bar{B}) = - f \frac{d}{dx} \log \chi_m(x; \bar{A}, \bar{B}) - \frac{1}{2} f',
\end{equation}
where $\chi_m(x;\bar{A}, \bar{B})$ may be one of the functions (\ref{eq:chi}) with $A$ and $B$ replaced by $\bar{A}$ and $\bar{B}$ or, in other words, $\Delta_1$ and $\Delta_2$ becoming $\bar{\Delta}_1$ and $\bar{\Delta}_2$, defined in (\ref{eq:Delta-bar}). The results read
\begin{equation}
\begin{split}
  W^{(m)}_{\alpha}(x; \bar{A}, \bar{B}) &= - \frac{1}{2} [(1+\alpha) \Delta_1  + (1-\alpha) \Delta_2 + \alpha] 
       \sec x \\
  &\quad - \frac{1}{2} [(1+\alpha) \Delta_1 - (1-\alpha) \Delta_2 + 1] \tan x - f \frac{p'_m}{p_m} \\
  & \qquad \text{for type I}, \\
  W^{(m)}_{\alpha}(x; \bar{A}, \bar{B}) &= - \frac{1}{2} [-(1+\alpha) \Delta_1 - (1-\alpha) \Delta_2 + \alpha] 
       \sec x \\
  &\quad - \frac{1}{2} [-(1+\alpha) \Delta_1 + (1-\alpha) \Delta_2 + 1] \tan x - f \frac{p'_m}{p_m} \\
  & \qquad \text{for type II}, \\ 
  W^{(m)}_{\alpha}(x; \bar{A}, \bar{B}) &= - \frac{1}{2} [(1+\alpha) \Delta_1 - (1-\alpha) \Delta_2 + \alpha] 
      \sec x \\
   &\quad - \frac{1}{2} [(1+\alpha) \Delta_1 + (1-\alpha) \Delta_2 + 1] \tan x - f \frac{p'_m}{p_m} \\
   & \qquad \text{for type III}, 
\end{split}
\end{equation}
the corresponding $p_m = p_m^{(\Delta_1,\Delta_2)}(t)$ being defined in (\ref{eq:p_m}) and a dash denoting as before a derivative with respect to $x$. The associated energies ${\cal E}_m(\bar{A}, \bar{B})$ are given in Eq.~(\ref{eq:cal-E}) with $A$, $B$ replaced by $\bar{A}$, $\bar{B}$, and consequently $\Delta_1$, $\Delta_2$ changed into $\bar{\Delta}_1$, $\bar{\Delta_2}$, given in (\ref{eq:Delta-bar}).\par
%
%
As it can be checked by using the differential equation satisfied by Jacobi polynomials \cite{olver}, in the three cases we can write
\begin{equation}
\begin{split}
  & \left[W^{(m)}_{\alpha}(x;\bar{A},\bar{B})\right]^2 - f W^{(m)'}_{\alpha}(x;\bar{A},\bar{B}) + {\cal E}_m(\bar{A},    
      \bar{B})) = V(x;\bar{A},\bar{B}), \\
  & \left[W^{(m)}_{\alpha}(x;\bar{A},\bar{B})\right]^2 + f W^{(m)'}_{\alpha}(x;\bar{A},\bar{B}) + {\cal E}_m(\bar{A},    
      \bar{B})) = V^{(m)}_{\rm ext}(x;A,B) + \gamma.
\end{split}
\end{equation}
\par
%
%
In the two isospectral cases I and II, the final potential $V^{(m)}_{\rm ext}(x;A,B) + \gamma$ satisfies a deformed shape invariance property similar to that of the starting potential $V(x;\bar{A},\bar{B})$. Let us indeed consider a superpotential
\begin{equation}
  W^{(m)}_{\rm ext}(x;A,B) = - f \frac{d}{dx} \log \psi_0^{(\rm ext)}(x;A,B) - \frac{1}{2} f'. \label{eq:W-ext}
\end{equation}
with
\begin{equation}
  \psi_0^{(\rm ext)}(x;A,B) \propto \frac{\psi_0^{(\alpha)}(x;A,B)}{p_m^{(\Delta_1,\Delta_2)}(t)} Q_0^{(m)}(t;A,B)
  \label{eq:psi-ext}
\end{equation}
and
\begin{equation}
  Q_0^{(m)}(t;A,B) = - (\Delta_1+1) p_m^{(\Delta_1,\Delta_2)}(t) - \frac{1}{2}(m+\Delta_2-\Delta_1-1)
  (1-t) p_{m-1}^{(\Delta_1-1,\Delta_2+1)}(t),
\end{equation}
or
\begin{equation}
  Q_0^{(m)}(t;A,B) = (\Delta_2+1) p_m^{(\Delta_1,\Delta_2)}(t) - \frac{1}{2}(m+\Delta_1-\Delta_2-1)
  (1+t) p_{m-1}^{(\Delta_1+1,\Delta_2-1)}(t)
\end{equation}
for type I or II, respectively. On using the definition of $p_m^{(\Delta_1,\Delta_2)}(t)$, given in (\ref{eq:p_m}), and its expansion into powers \cite{olver}, it is straightforward to rewrite $Q_0^{(m)}(t;A,B)$ as
\begin{equation}
  Q_0^{(m)}(t;A,B) = (m-\Delta_1-1) p_m^{(\Delta_1+1,\Delta_2+1)}(t) \qquad \text{for type I},  \label{eq:Q-0-1}
\end{equation}
and
\begin{equation}
  Q_0^{(m)}(t;A,B) = - (m-\Delta_2-1) p_m^{(\Delta_1+1,\Delta_2+1)}(t) \qquad \text{for type II}.  \label{eq:Q-0-2}
\end{equation}
Combining Eqs. (\ref{eq:W-ext}) and (\ref{eq:psi-ext}) with (\ref{eq:Q-0-1}) and (\ref{eq:Q-0-2}) yields
\begin{equation}
  W^{(m)}_{\rm ext}(x;A,B) = \lambda \tan x + \mu \sec x - f \left(\frac{p_m^{(\Delta_1+1,\Delta_2+1)'}}
  {p_m^{(\Delta_1+1,\Delta_2+1)}} - \frac{p_m^{(\Delta_1,\Delta_2)'}}{p_m^{(\Delta_1,\Delta_2)}}\right)
\end{equation}
for both types. It is then straightforward to show that the partner of $V^{(m)}_{\rm ext}(x;A,B)$ is given by
\begin{equation}
  V^{(m)}_{\rm ext}(x;A,B) + 2f W^{(m)'}_{\rm ext}(x;A,B) = V^{(m)}_{\rm ext}(x;A^{(1)},B^{(1)}) + \gamma
\end{equation}
with $A^{(1)}$, $B^{(1)}$ considered in (\ref{eq:A1-B1}).\par
%
%
\section{A simple example}

\setcounter{equation}{0}

The simplest example of rationally-extended Scarf I potential with PDM corresponds to $m=1$ in (\ref{eq:V_0-V_1}) and may be obtained as a type I or II potential. The result reads
\begin{align}
  V^{(1)}_{\rm ext}(x;A,B) & = [A(A-1) + B^2] \sec^2 x - B(2A-1) \sec x \tan x \nonumber \\
  & \quad + 2 (1-\alpha^2) (1+\alpha \sin x) \left(\frac{\Delta_1+\Delta_2}{q^{(\Delta_1,\Delta_2)}(x)} -
      \frac{4\Delta_1\Delta_2(1+\alpha \sin x)}{[q^{(\Delta_1,\Delta_2)}(x)]^2}\right), \label{eq:pot-1} 
\end{align}
where
\begin{equation}
  q^{(\Delta_1,\Delta_2)}(x) = (1+\alpha)\Delta_1 + (1-\alpha)\Delta_2 + [(1+\alpha)\Delta_1 - (1-\alpha)\Delta_2]
  \sin x.
\end{equation}
\par
%
%
Its spectrum is given in Eq.~(\ref{eq:E-ext-n}). Its wavefunctions, expressed in (\ref{eq:wf-ext-1}) and (\ref{eq:wf-ext-1a}) or (\ref{eq:wf-ext-1b}), may also be written in a simpler way in terms of the $(n+1)$th-degree $X_1$-Jacobi EOP's $\hat{P}^{(\Delta_1,\Delta_2)}_{n+1}(t)$, $n=0$, 1, 2, \ldots, defined in \cite{gomez09}. This is done by applying the PCT considered above on a known result for a rationally-extended Scarf I potential in a constant-mass background \cite{cq09a} The result 
reads
\begin{align}
  \psi^{(\rm ext}_n(x;A,B) &= {\cal N}^{(\rm ext)}_{n,A,B} f^{-\frac{1}{2}(\Delta_1+\Delta_2+2)} 
      \frac{(1-\sin x)^{\frac{1}{2}(\Delta_1+\frac{1}{2})} (1+\sin x)^{\frac{1}{2}(\Delta_2+\frac{1}{2})}}
      {q^{(\Delta_1,\Delta_2)}(x)} \nonumber \\
   &\quad \times \hat{P}^{(\Delta_1,\Delta_2)}_{n+1}(t)
\end{align}
with $t=(\alpha+\sin x)/(1+\alpha \sin x)$ and
\begin{align}
  {\cal N}^{(\rm ext)}_{n,A,B} &= \frac{\Delta_2-\Delta_1}{2^{(\Delta_1+\Delta_2+1)/2}} \nonumber \\
  &\quad\times \left[(1-\alpha)^{\Delta_1+1}
       (1+\alpha)^{\Delta_2+1} (\Delta_1+\Delta_2+2n+1) n! \Gamma(\Delta_1+\Delta_2+n+1)\right]^{1/2}
       \nonumber \\
  &\quad \times \left[(\Delta_1+n+1) (\Delta_2+n+1) \Gamma(\Delta_1+n) \Gamma(\Delta_2+n)\right]^{-1/2}.
\end{align}
\par
%
%
{}For the lowest $n$ values, for instance, we obtain
\begin{equation}
\begin{split}
  \hat{P}^{(\Delta_1,\Delta_2)}_1(t) &= \frac{1}{2(\Delta_2-\Delta_1)(1+\alpha\sin x)} [(1+\alpha)(\Delta_1+1)
      (1+\sin x) \\
  &\quad + (1-\alpha)(\Delta_2+1)(1-\sin x)], \\
  \hat{P}^{(\Delta_1,\Delta_2)}_2(t) &= \frac{1}{4(\Delta_2-\Delta_1)(1+\alpha\sin x)^2} [(1+\alpha)^2 \Delta_1
      (\Delta_1+2) (1+\sin x)^2 \\
  &\quad -(1-\alpha)^2 \Delta_2(\Delta_2+2) (1-\sin x)^2], \\
  \hat{P}^{(\Delta_1,\Delta_2)}_3(t) &= \frac{1}{16(\Delta_2-\Delta_1)(1+\alpha\sin x)^3} [(1+\alpha)^3 \Delta_1
      (\Delta_1+1)(\Delta_1+3)(1+\sin x)^3 \\
  &\quad - (1+\alpha)^2 (1-\alpha) \Delta_1(\Delta_1+3)(\Delta_2+3)(1+\sin x)^2(1-\sin x) \\
  &\quad - (1+\alpha)(1-\alpha)^2 \Delta_2(\Delta_1+3)(\Delta_2+3) (1+\sin x)(1-\sin x)^2 \\
  &\quad + (1-\alpha)^3 \Delta_2(\Delta_2+1)(\Delta_2+3) (1-\sin x)^3].
\end{split}
\end{equation}
\par
%
%
\section{Conclusion}

In the present paper, we have shown that the problem of the Scarf I potential in a $m(\alpha:x) = (1+\alpha \sin x)^{-2}$ background can be easily solved by mapping the corresponding Schr\"odinger equation onto that of the Scarf I potential with constant mass by means of a PCT. In addition, we have proved that the well-known shape invariance of the Scarf I potential in SUSYQM gives rise to a deformed shape invariance property for the considered PDM problem in deformed SUSYQM.\par
%
%
We have also taken advantage of the knowledge of rational extensions of the Scarf I potential connected with $X_m$-Jacobi EOPs of type I, II, and III to transform them into rational extensions for the corresponding PDM problem. These results have been analyzed in the deformed SUSYQM framework and the extended Scarf I potentials of type I and II have been shown to be endowed with a deformed shape invariance property.\par
%
%
Considering multi-indexed rational extensions and corresponding orthogonal polynomials would be a very interesting topic for future investigation.\par
%
%
\section*{Data availability statement}

No new data were created or analyzed in this study.\par
%
%
\section*{Acknowledgment}

This work was supported by the Fonds de la Recherche Scientifique - FNRS under Grant Number  4.45.10.08.\par
%
%
\section*{Author contribution}

Conceptualization (equal). Formal analysis (equal). Investigation (equal). Methodology (equal).\par
%
%
\section*{Appendix}

\renewcommand{\theequation}{A.\arabic{equation}}
\setcounter{equation}{0}

The purpose of this appendix is to determine the explicit form of the parameters $\bar{A}$ and $\bar{B}$ of the first partner potential $V_0(x)$ considered in (\ref{eq:V_0-V_1}), as well as that of the parameters $A^{(1)}$ and $B^{(1)}$ of the partner potential $V_1(x)$ in (\ref{eq:V_1}).\par
%
%
Since $\bar{A}$ and $\bar{B}$ correspond to the parameters $\bar{A}'$ and $\bar{B}'$ defined in (\ref{eq:g}) or, in other words to $(\bar{\Delta}_1, \bar{\Delta}_2)$ given in (\ref{eq:Delta-bar}), they satisfy the equations
\begin{equation}
\begin{split}
  &\left[\frac{1}{4} + \frac{(\bar{A}-\bar{B})(\bar{A}-\bar{B}-1)}{(1+\alpha)^2}\right]^{1/2} =
      \left[\frac{1}{4} + \frac{(A-B)(A-B-1)}{(1+\alpha)^2}\right]^{1/2} +1, \\
  &\left[\frac{1}{4} + \frac{(\bar{A}+\bar{B})(\bar{A}+\bar{B}-1)}{(1-\alpha)^2}\right]^{1/2} =
      \left[\frac{1}{4} + \frac{(A+B)(A+B-1)}{(1-\alpha)^2}\right]^{1/2} -1  
\end{split}. \label{eq:A1}
\end{equation}
for type I potentials. Equation (\ref{eq:A1}) leads to the relations
\begin{equation}
\begin{split}
  &\bar{A} - \bar{B} = \frac{1}{2}\left\{1 + \left[(2A-2B-1)^2 + 4(1+\alpha)^2 (1+2\Delta_1)\right]^{1/2}\right\}, \\
  &\bar{A} + \bar{B} = \frac{1}{2}\left\{1 + \left[(2A+2B-1)^2 + 4(1-\alpha)^2 (1-2\Delta_2)\right]^{1/2}\right\},
\end{split}. \label{eq:A2}
\end{equation}
from which we get
\begin{equation}
\begin{split}
  \bar{A} &= \frac{1}{2}\biggl\{1 + \frac{1}{2}\left[(2A-2B-1)^2 + 4(1+\alpha)^2 (1+2\Delta_1)\right]^{1/2} \\
      &\quad + \frac{1}{2} \left[(2A+2B-1)^2 + 4(1-\alpha)^2 (1-2\Delta_2)\right]^{1/2}\biggr\}, \\
  \bar{B} &= \frac{1}{4}\biggl\{- \left[(2A-2B-1)^2 + 4(1+\alpha)^2 (1+2\Delta_1)\right]^{1/2} \\
  &\quad + \left[(2A+2B-1)^2 + 4(1-\alpha)^2 (1-2\Delta_2)\right]^{1/2}\biggr\}.
\end{split}
\end{equation}
\par
%
%
On proceeding in a similar way for type II and III potentials, we get
\begin{equation}
\begin{split}
  \bar{A} &= \frac{1}{2}\biggl\{1 + \frac{1}{2}\left[(2A-2B-1)^2 + 4(1+\alpha)^2 (1-2\Delta_1)\right]^{1/2} \\
      &\quad + \frac{1}{2} \left[(2A+2B-1)^2 + 4(1-\alpha)^2 (1+2\Delta_2)\right]^{1/2}\biggr\}, \\
  \bar{B} &= \frac{1}{4}\biggl\{- \left[(2A-2B-1)^2 + 4(1+\alpha)^2 (1-2\Delta_1)\right]^{1/2} \\
  &\quad + \left[(2A+2B-1)^2 + 4(1-\alpha)^2 (1+2\Delta_2)\right]^{1/2}\biggr\},
\end{split}
\end{equation}
and
\begin{equation}
\begin{split}
  \bar{A} &= \frac{1}{2}\biggl\{1 + \frac{1}{2}\left[(2A-2B-1)^2 + 4(1+\alpha)^2 (1+2\Delta_1)\right]^{1/2} \\
      &\quad + \frac{1}{2} \left[(2A+2B-1)^2 + 4(1-\alpha)^2 (1+2\Delta_2)\right]^{1/2}\biggr\}, \\
  \bar{B} &= \frac{1}{4}\biggl\{- \left[(2A-2B-1)^2 + 4(1+\alpha)^2 (1+2\Delta_1)\right]^{1/2} \\
  &\quad + \left[(2A+2B-1)^2 + 4(1-\alpha)^2 (1+2\Delta_2)\right]^{1/2}\biggr\},
\end{split}. \label{eq:A5}
\end{equation}
respectively. It is worth noting that all quantities under a square root in (\ref{eq:A2})--(\ref{eq:A5}) can be easily checked to be positive, so that the parameters $\bar{A}$ and $\bar{B}$ are real.\par
%
%
On considering next the problem of $A^{(1)}$ and $B^{(1)}$ appearing in (\ref{eq:V_1}), we note that this time
\begin{equation}
\begin{split}
  \left[\frac{1}{4} + \frac{(A^{(1)}-B^{(1)})(A^{(1)}-B^{(1)}-1)}{(1+\alpha)^2}\right]^{1/2}  
      &= \left[\frac{1}{4} + \frac{(A^{(0)}-B^{(0)})(A^{(0)}-B^{(0)}-1)}{(1+\alpha)^2}\right]^{1/2} \\
  &\quad +1, \\
  \left[\frac{1}{4} + \frac{(A^{(1)}+B^{(1)})(A^{(1)}+B^{(1)}-1)}{(1-\alpha)^2}\right]^{1/2}  
    &= \left[\frac{1}{4} + \frac{(A^{(0)}+B^{(0)})(A^{(0)}+B^{(0)}-1)}{(1-\alpha)^2}\right]^{1/2} \\
  &\quad +1. 
\end{split}
\end{equation}
Hence, by proceeding as above, we obtain
\begin{equation}
\begin{split}
  A^{(1)} &= \frac{1}{2} \biggl\{1 + \frac{1}{2}\left[\left(2A^{(0)}-2B^{(0)}-1\right)^2 + 4(1+\alpha)^2 (1+
      2 \Delta_1)\right]^{1/2} \\
  &\quad + \frac{1}{2}\left[\left(2A^{(0)}+2B^{(0)}-1\right)^2 + 4(1-\alpha)^2 (1+2\Delta_2)\right]^{1/2} \biggr\},\\
  B^{(1)} &= \frac{1}{4} \biggl\{-\left[\left(2A^{(0)}-2B^{(0)}-1\right)^2 + 4(1+\alpha)^2 (1+
      2 \Delta_1)\right]^{1/2} \\
  &\quad + \left[\left(2A^{(0)}+2B^{(0)}-1\right)^2 + 4(1-\alpha)^2 (1+2\Delta_2)\right]^{1/2} \biggr\}.
\end{split}
\end{equation}
\par
%
%

\end{document}